\documentclass[useAMS,usenatbib]{mn2e}
\usepackage{psfig,subfigure}

\title[Monte Carlo Photoionization Simulations of Diffuse 
Ionized Gas]
 {Monte Carlo Photoionization Simulations of Diffuse 
Ionized Gas}

\author[Wood \& Mathis]
 {Kenneth Wood$^1$ and John S. Mathis$^2$ \\
 $^1$School of Physics \& Astronomy, University of St. Andrews,\\
North Haugh, St Andrews, KY16 9SS, Scotland;
kw25@st-andrews.ac.uk\\
 $^2$Astronomy Department, University of Wisconsin, \\
475 N. Charter Street, Madison, WI 53706; mathis@astro.wisc.edu\\
}
\date{Released 2002 Xxxxx XX}
\pagerange{\pageref{firstpage}--\pageref{lastpage}} \pubyear{2002}

\def\LaTeX{L\kern-.36em\raise.3ex\hbox{a}\kern-.15em
 T\kern-.1667em\lower.7ex\hbox{E}\kern-.125emX}

\begin{document}

\label{firstpage}

\maketitle

\begin{abstract}

We demonstrate that the observed increase of some nebular line ratios 
with height above the midplane in the diffuse ionized gas (DIG) in 
the Milky Way and other galaxies 
is a natural consequence of the progressive hardening 
of the radiation field far from the midplane ionizing sources.  
To obtain increasing temperatures and line ratios away from the midplane, 
our photoionization simulations of a multi-component 
interstellar medium do not require as much additional 
heating (over and above that from photoionization) as previous 
studies that employed one dimensional, spherically averaged models.  

Radiation leaking into the DIG from density 
bounded H~{\sc ii} regions is generally harder in the H-ionizing continuum 
and has its He-ionizing photons suppressed compared to the ionizing source 
of the H~{\sc ii} region.  In line with other recent investigations, 
we find that such 
leaky H~{\sc ii} region models can provide elevated temperatures and 
line ratios, and a lower He$^+$ fraction in the DIG.  
For a composite model representing the relative spectral types of O stars 
in the solar neighbourhood, we find that additional heating less than 
$10^{-26} n_{\rm e}$~ergs/s/cm$^3$ can reproduce 
the observed elevated line ratios in the DIG.  This additional 
heating is considerably 
less than previous estimates due to the natural hardening of 
the radiation field reaching large heights in our simulations.

\end{abstract}


\section{Introduction}

The presence of extended layers of diffuse ionized gas (DIG) 
in the Milky Way and other galaxies is inferred 
from faint emission in H$\alpha$ and other nebular lines 
(e.g., Reynolds 1995; Rand 1998; Hoopes, Walterbos, \& Greenawalt 1996; 
Domgoergen \& Dettmar 1997; Otte \& Dettmar 1999; Hoopes \& Walterbos 2003; 
Wang, Heckman, \& Lehnert 1998; Otte et al. 2001, 2002).  
The ionization 
source for the DIG is believed to be O and B stars close 
to the midplane of galaxies.  Photoionization models using OB stars can 
reproduce the gross features of the DIG ionization structure 
(e.g., Miller \& Cox 1993; Dove \& Shull 1994).  If extra heating is 
included, in addition to that from photoionization by O stars, 1D models can 
reproduce the DIG spectrum (Mathis 1986; 
Domgoergen \& Mathis 1994; Mathis 2000; Sembach et al. 2000).  
The main problem with O stars as the ionization source of the DIG 
is that for a smooth density distribution the mean free path 
of Lyman continuum photons is very small, $\sim 0.1$~pc for 
$n({\rm H}^0)=1\, {\rm cm}^{-3}$. Thus, it is difficult for photons to 
traverse the large distances to ionize the extraplanar gas which has a 
scaleheight of around 1~kpc in the Milky Way 
(Haffner, Reynolds, \& Tufte 1999).  However, three dimensional 
photoionization models show that a two component (Wood \& Loeb 2000) or 
fractal ISM (Ciardi, Bianchi, \& Ferrara 2002) can provide low density paths 
allowing Lyman continuum photons to reach large $|z|$ heights above their 
midplane sources.

Measurements of emission lines in addition to H$\alpha$ provide probes of 
the abundances, temperatures, and densities in 
the DIG.  Some of the most studied lines in the Milky Way and other 
galaxies are [N~II] $\lambda 6584$, [S~II] $\lambda 6717$, 
[O~III] $\lambda 5007$, [O~II] $\lambda 3727$ doublet, 
and [O~I] $\lambda 6300$.  
Observations of the He~I $\lambda 5876$ line provide information on the 
spectrum of the ionizing sources (Reynolds \& Tufte 1995).  Hereafter 
the above lines will simply be referred to as [N~II], [S~II], [O~III], 
[O~II], [O~I], and He~I.  
Some differences in the observed line ratios in the DIG compared to 
traditional H~{\sc ii} regions are: [S~II]/H$\alpha$ and [N~II]/H$\alpha$ 
ratios increase with height above the plane; [S~II]/[N~II] is quite 
uniform with height (Haffner et al. 1999) and with latitude (Rand 1997); 
He is observed to be underionized with respect to H 
(Reynolds \& Tufte 1995; Heiles et al. 1996).

Most models for the DIG 
employ smooth, one dimensional density distributions, and predict 
volume averages of the line strengths and ratios (e.g., Mathis 1986; 
Domgoergen \& Mathis 1994; Collins \& Rand 2001; Sembach et al. 2000; 
Mathis 2000).  These models generally fail 
to reproduce the observed line ratios that increase with $|z|$ above the 
midplane.  Haffner et al. (1999) and Reynolds, Haffner, \& Tufte (1999) 
showed that the observed [S~II]/H$\alpha$ and [N~II]/H$\alpha$ line ratios 
may be explained if the gas temperature increases with height above the 
midplane.  Including additional heating over and above that of pure 
photoionization can reconcile the one dimensional models with 
observations (Reynolds et al. 1999; Mathis 2000).  Additional 
heating may plausibly arise from photoelectric heating from dust 
(Reynolds \& Cox 1992), dissipation of turbulence (Slavin et al. 1993; 
Minter \& Spangler 1997), and shocks (Raymond 1992).  
However, it is well known that in photoionized 
H~{\sc ii} regions the radiation field hardens towards the edge of the 
Str{\"o}mgren sphere resulting in the highest temperatures occurring at the 
largest distances from the ionizing source.  This arises because low
energy photons have relatively short mean free paths and are absorbed 
close to the source.  Higher energy photons have longer mean free paths, 
travel farther, and deposit more energy per photon at large distances from 
the source, giving rise to the increasing temperature away from the source 
(Osterbrock 1989).  

Such a scenario as described above is almost certainly occurring in the 
DIG, with the ionizing spectra penetrating to large $|z|$ above the plane 
being significantly harder than the source spectra, naturally producing 
a temperature profile that increases with $|z|$.  Models presented by 
Bland-Hawthorn, Freeman, \& Quinn (1997), Wang et al. (1998), and 
Rand (1998) take this radiation transfer effect into account by introducing a 
hardening of the radiation field with $|z|$.  
In these plane parallel models, the hardening of the radiation field leads 
to increasing temperatures at large distances from the illuminated face and 
corresponding increases in [S~II]/H$\alpha$ and [N~II]/H$\alpha$.  
Mathis (2000) also mentions this effect, 
stating that line ratios for lines of sight that pierce the outer edges of 
spherical models may reproduce the observations.  

Radiation leaking into the ISM from traditional H~{\sc ii} regions may 
also be hardened compared to the source spectrum.  Hoppes \& Walterbos 
(2003) investigated photoionization by photons from 
leaky H~{\sc ii} regions finding 
that the hardened spectrum leads to elevated temperatures and increased 
line ratios when compared to models that do not envoke hardening of the 
source spectrum.  In addition, leaky H~{\sc ii} region models lead to 
a suppression of He-ionizing photons ($h\nu > 24.6$~eV) and a 
corresponding decrease in ionization stages such as He$^+$ and N$^{2+}$.

In this paper we present Monte Carlo photoionization  
models for a multi-component ISM.   In what follows we describe 
the photoionization code, adopted ISM density structure, and spectra for 
the ionizing sources.  Due to the very large parameter space, we restrict 
this paper to two dimensional models of a single source ionizing a 
multi component, stratified ISM.  Note that although our models are 
for 2D systems, our simulations are run on 3D density grid.  
A future paper will present models for 3D geometries and illuminations, 
investigating the role of diffuse ionizing radiation, 
3D ionization and temperature structures, and the resulting intensity maps.  
Some specific issues we address in this paper are: increased 
temperatures with increasing $|z|$ above the midplane; problems arising 
with fitting [S~II] emission due to undetermined dielectronic recombination 
rates; predictions of excess [O~I] emission compared with observations; the 
ionization of He within the DIG.

\section{Models}

Our simulations use a 3D Monte Carlo photoionization code, with inputs 
being the ISM density structure, and the locations and ionizing spectra 
of point sources within the simulation grid.  These ingredients are 
briefly discussed below.

\subsection{Photoionization Models}

The photoionization simulations are performed using the 3D Monte Carlo 
code of Wood, Mathis, \& Ercolano (2004).  The code simulates 
photoionization due to multiple point or extended sources and discretizes 
the density onto a 3D linear Cartesian grid.  For all simulations in 
this paper the grid comprises $65^3$ cells.  The code performs well 
compared to traditional 1D codes and other independently developed 
Monte Carlo codes (Ercolano et al. 2003; Och, Lucy, \& Rosa 1998).  
We do not consider the effects of shocks or ionization fronts.   
In our simulations there is no C$^0$ or S$^0$ because we assume C and S 
are fully ionized by the ambient interstellar radiation field.  There 
are no ionizing photons with energies above 54.4~eV, so there is no 
${\rm He}^{2+}$, consistent with observations of almost all H~{\sc ii} 
regions.  For further details of the code and comparisons with other 
codes see Wood et al. (2004).

For most simulations, 
we adopt the following abundances by number relative to H: 
${\rm He/H}=0.1$, ${\rm C/H} = 140$~ppm, 
${\rm N/H} = 75$~ppm, ${\rm O/H} = 319$~ppm, 
${\rm Ne/H} = 117$~ppm, and ${\rm S/H} = 18.6$~ppm.  
With the exception of S/H, these abundances were used by Mathis (2000) in 
his photoionization models of the local DIG.  We found that the solar S/H 
(Anders \& Grevesse 1989) provides a good match to the observations.  The 
lower value, S/H = 13 ppm, used by Mathis (2000) was probably due to 
different dielectronic recombination rates in his photoionization 
code (see discussion in \S~3.5.4).  
We also investigate the effects of different abundances and 
perform some simulations using abundances appropriate for the Perseus arm 
(Mathis 2000).  
We do not consider cooling from collisionally excited lines from elements 
other than those listed here, so our temperatures may be slightly hotter 
than 1D models that include more elements (e.g., Sembach et al. 2000).

Currently dielectronic recombination rates have not been calculated for 
third and fourth row elements, so photoionization modeling of sulfur lines 
are subject to uncertainties due to unknown atomic data (see 
Ali et al. 1991).  
We use the total recombination rates for S$^+$ and 
S$^{2+}$ from Nahar (2000).  For S$^{3+}$ we use the radiative 
recombination rates from Verner \& Ferland (1996) and the average 
dielectronic rate of $2.5\times 10^{-11}$ from Ali et al. (1991).  We 
discuss the issues related to modeling [S~II] lines in \S~3.5.4.

\subsection{ISM Density Structure}

The ISM is observed to comprise several components of (warm and cold) 
neutral and (warm and hot) ionized gas.  
Unlike many recent photoionization models of the DIG, Monte Carlo 
techniques are not restricted to 1D averaged models.  
For our smooth density models, we use the two component density from
Miller \& Cox (1993) 
\begin{equation}
n({\rm H}) = 0.1\exp(-|z|/0.3) + 0.025\exp(-|z|/0.9)\; ,
\end{equation}
where the number densities are per cm$^3$ and the distances are in kpc.  
This represents the concentrated neutral layer and the extended ionized 
layer.  
In their model for the DIG, Miller \& Cox used the above smooth density and 
included an approximation for absorption by dense clouds using a model 
that reproduces the statistics of clouds in the ISM.  
Although this smooth density is lower than the average density inferred 
for H$^0$ and H$^+$, when implemented in their ``standard cloud'' model 
using the known ionizing sources in the solar neighbourhood, it reproduces 
the average emission and dispersion measures observed in the Milky Way.  

An obvious criticism of the Miller \& Cox model is that the density they 
used is smaller than that inferred for the H~I in the Galaxy (e.g., 
Dickey \& Lockman 1990).  However, they 
did use the known distribution and ionizing luminosity of O stars in the 
Solar neighbourhood, and found that the above density allowed for the gas 
to be ionized at large $z$ along with reproducing many of the observations 
of the DIG.  In reality the ISM is clumped on a large range of sizescales, 
so the Miller \& Cox density provides an estimate of the smooth component 
required to allow ionizing photons to penetrate to large $|z|$.  Therefore, 
such a density is a good starting point for 3D models that incorporate 
smooth and clumpy components: the smooth component should be close to that 
used by Miller \& Cox.

\subsection{Ionizing Sources and Spectra}

We consider single sources within multi-component 
ISM density distributions.  The ionizing spectra are taken from 
the WM-basic model atmosphere library (Pauldrach et al. 2001; 
Sternberg, Hoffmann, \& Pauldrach 2004) which provides 
model atmospheres and emergent spectra including the effects of non-LTE 
line-blanketing and stellar winds.  We did not run the WM-basic models, 
but used spectra from the library generated by Sternberg et al. (2004).  
The spectra used were for model atmospheres with solar abundances, 
and gravities and temperatures in the range $3.6\le \log g \le 4.0$ and 
30000~K $\le T_\star \le$ 50000~K.

\section{Standard Model}

\begin{figure}
\psfig{figure=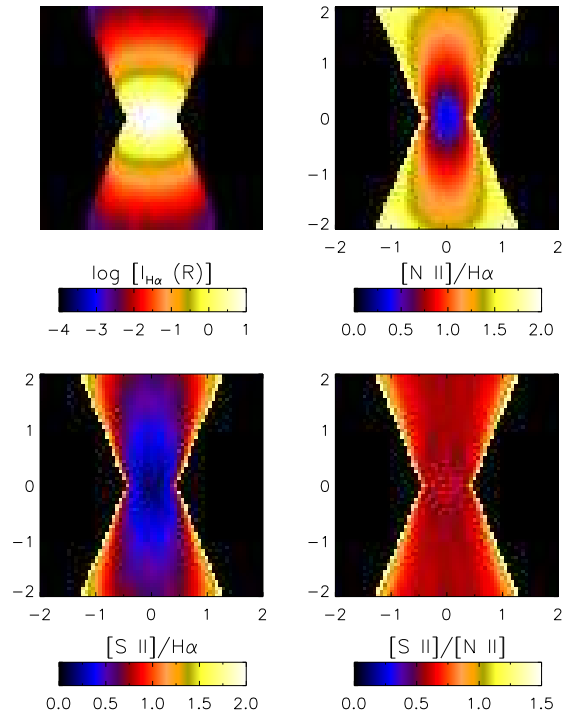,width=3.5truein,angle=0,height=4.08truein}
\caption[]{Total projected intensity and line ratio maps for 
H$\alpha$, 
[N~II]/H$\alpha$, [S~II]/H$\alpha$, and [S~II]/[N~II].  Axes are labeled in 
kpc.  The 
anti-correlation of [S~II] and [N~II] with H$\alpha$ and the rise of 
[N~II]/H$\alpha$, [S~II]/H$\alpha$ towards the edge of the ionized zone 
is evident. [S~II]/[N~II] is fairly constant within the ionized region, 
but rises towards the outer edges.}
\end{figure}

\begin{figure}
\psfig{figure=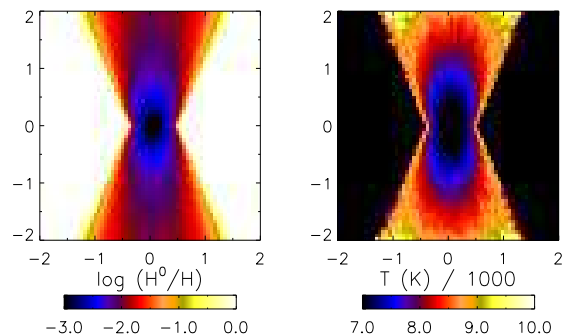,width=3.0truein,angle=0,height=1.80truein}
\caption[]{Two dimensional slice through the centre of the simulation 
grid showing the neutral H fraction and temperature.  The increase in 
temperature towards the edge of the ionized region due to hardening 
of the radiation field is clearly seen.}
\end{figure}

Our standard single source model has the two component, 
smooth density structure of Eq.~1, an ionizing luminosity 
$Q({\rm H}^0) = 6 \times 10^{49} {\rm s}^{-1}$, and an ionizing spectrum 
for a WM-basic model atmosphere with $T_\star = 40000$~K and $\log g =3.75$.  
The output of the photoionization simulation is the 3D temperature and 
ionization structure for the elements that we track.  Using the output 
temperature and ionization grids, we then calculate emissivity grids 
for various lines and form maps of the intensity and line ratios.  
The results of our standard simulation are shown in Figs.~1 -- 5.  

In forming the emissivity grid we do not include cells that have 
${\rm H}^0/{\rm H}>0.25$.  Emissivities for cells in our simulations that 
have higher neutral fractions are unreliable due to the resolution of our 
grid.  Also, because we do not include the dynamics of ionization fronts, 
emissivities from such cells are not correct since the physics of the 
ionization front is not 
included in our simulations (see Wood et al. 2004 for further discussion).  

Numerical noise in the Monte Carlo simulations is most easily seen in the 
figures showing cuts across the line ratio maps in Fig.~3 and the 
[N~II]/[S~II] cut in Fig.~5.  The ``jaggedness'' of the lines in these 
figures can be taken as an estimate of the numerical noise in forming the 
various line ratios.
Noise is also seen in the [S~II]/H$\alpha$ 
vs H$\alpha$ and [S~II]/H$\alpha$ vs H$\alpha$ scatter plots in Fig.~5, 
with some points deviating from the clear anti-correlation.  Although 
these discrepant points appear to give a large spread in an otherwise 
tight anti-correlation, they represent only a few percent of all 
points plotted and are generally towards the edge of the ionized zone 
where our photon statistics are poorest.

Figure~1 shows maps of the H$\alpha$ intensity (in Rayleighs), 
and the [N~II]/H$\alpha$, 
[S~II]/H$\alpha$, and [S~II]/[N~II] line ratios.  In calculating the 
H$\alpha$ intensity we have assumed that all the emission from the 
simulation grid is at a distance of 2.5~kpc, roughly the distance to the 
Perseus arm (Haffner et al. 1999).  
This figure immediately 
shows the anticipated increase in [N~II]/H$\alpha$ and [S~II]/H$\alpha$ 
with increasing distance from the central ionizing source.  The increasing 
line ratios arise from the hardening of the radiation field and subsequent 
increasing temperatures at large distances from the source.  

For this simulation [S~II]/[N~II]~$\sim 0.6$, 
similar to that seen in the local DIG (Haffner et al. 1999), and rises rapidly 
at the outermost edges of the ionized region.  The rapid rise is explained 
by the relative ionization states of the elements.  Close to the edge of 
the ionized region, N is mostly N$^+$ and then transitions rapidly to N$^0$, 
while S is transitioning from S$^{2+}$ to S$^+$ (there is no S$^0$ in 
our simulations).  Therefore, S$^+$/S increases and N$^+$/N decreases at the 
edge of the ionized zone leading to the large [S~II]/[N~II] line ratios 
at the outer edges of the line ratio map (see also Bland-Hawthorn et al. 
1997; Fig.~9).  
The rapid increase in line ratios from the interface 
in our simulations is generally not observed in the DIG and may 
reflect a shortcoming of our pure photoionization models for this interface.  

Figure~2 shows a slice through the grid in the $x$-$z$ plane showing the 
ionization of H and the 2D temperature structure.  The ionized volume is 
extended perpendicular to the plane due to the increased density in 
the midplane.  This resembles the ``Str{\"o}mgren volumes '' seen in other 
2D photoionization simulations.  The increasing 
temperature with distance from the source is clearly seen in this figure.

Figure~3 shows line ratios as a function of $z$ for vertical cuts at 
$x=0$ and 0.4~kpc from the centre of the intensity maps in Fig.~1.  
The rise of [N~II]/H$\alpha$ 
and [S~II]/H$\alpha$ with increasing $|z|$ is apparent as discussed above.  
[O~III]/H$\alpha$ decreases with height above the plane, 
while [O~I]/H$\alpha$ and [O~II]/H$\alpha$ increase with increasing $|z|$.  
The He~I/H$\alpha$ ratio decreases with $|z|$ because the radiation 
reaching these regions has had some of its He-ionizing photons absorbed at 
lower $|z|$.  For low density gas at 10$^4$~K, 
${\rm He~I}/{\rm H}\alpha=0.5\,{\rm He}^+/{\rm H}^+$ (Osterbrock 1989), 
so the He~I/H$\alpha$ ratio probes the helium ionization  and 
in turn the ionizing spectrum of the DIG (Reynolds \& Tufte 1995).  
Figure~4 shows the incident spectrum and the spectrum that 
reaches $|z|=2$~kpc, clearly showing the hardening of the radiation field 
and suppression of the He-ionizing photons.

\begin{figure}
\psfig{figure=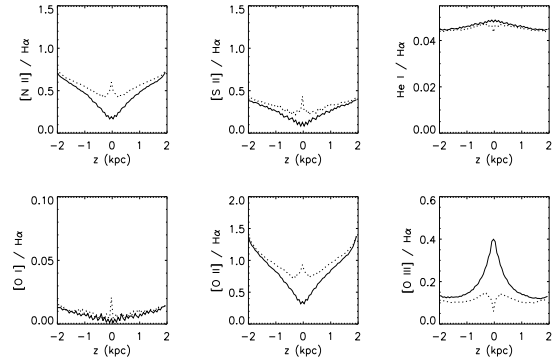,width=3.0truein,angle=0,height=2.0truein}
\caption[]{Vertical cuts at $x=0$~kpc (solid) 
and $x=0.4$~kpc (dots) showing the variation of line ratios with $z$.   }
\end{figure}

\begin{figure}
\psfig{figure=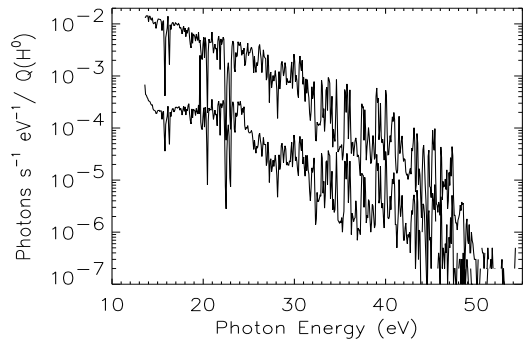,width=3.0truein,angle=0,height=2.0truein}
\caption[]{Incident source spectrum (upper) and spectrum reaching 
$|z|=2$~kpc (lower).  Compared to the source spectrum, 
the spectrum at large $|z|$ is harder in the 
H-ionizing continuum and has had its He-ionizing photons suppressed}
\end{figure}

\begin{figure}
\psfig{figure=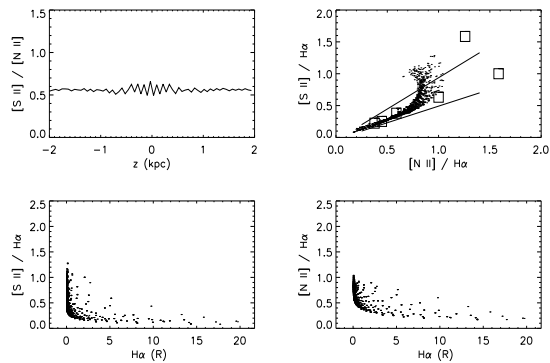,width=3.0truein,angle=0,height=2.0truein}
\caption[]{Vertical cut at $x=0$~kpc showing the variation of 
[S~II]/[N~II] with $z$.  The scatter plots show the correlation of 
[S~II]/H$\alpha$ with [N~II]/H$\alpha$ and the anticorrelation of 
[S~II]/H$\alpha$ and [N~II]/H$\alpha$ with H$\alpha$.  In the 
[S~II]/H$\alpha$ --- [N~II]/H$\alpha$ plot, the solid lines mark the 
range of observations in the local DIG (Haffner et al. 1999) and the 
squares show results from NGC~891 (Rand 1998).}
\end{figure}

In Fig.~5 we show a vertical cut across 
the centre of the [S~II]/[N~II] line ratio map.  As in Fig.~1 we see 
[S~II]/[N~II]~$\sim 0.6$, similar to observations 
by Haffner et al. (1999).
Figure~5 also shows some scatter plots as another way of displaying 
the trends seen in the intensity and line ratio maps of Fig.~1.  These 
scatter plots display the values for all pixels in the line ratio images 
of Fig.~1.  
Such scatter plots (e.g., Haffner et al. 1999) show the correlation of 
[S~II]/H$\alpha$ with [N~II]/H$\alpha$ and the anticorrelation of 
[S~II]/H$\alpha$ and [N~II]/H$\alpha$ with H$\alpha$.  The solid lines 
in the [S~II]/H$\alpha$ --- [N~II]/H$\alpha$ scatter plot 
show the range of observations in the local DIG (Haffner et al. 1999) and 
the squares are those for NGC~891 (Rand 1998).  
The increase in [S~II]/[N~II] towards the edges of the ionized volume 
in our simulations is evident in the change of slope 
in the [S~II]/H$\alpha$ vs [N~II]/H$\alpha$ scatter plot.  Note that 
the dots are appearing to turn around at very large values of 
[S~II]/H$\alpha$, where the [N~II]/H$\alpha$ ratio 
is decreasing due to the ${\rm N}^{+}\rightarrow {\rm N}^0$ transition.  
Our grid resolution does not allow us to follow the decrease of 
[N~II]/H$\alpha$ smoothly to zero.  
The change of slope in the [S~II]/H$\alpha$ --- [N~II]/H$\alpha$ scatter 
plot is not present in data from the local DIG, 
but may be present in the NGC~891 data.  Again, 
either the DIG is fully ionized or our models do not correctly treat 
emission from the ionized/neutral interface.  
The variation 
of [S~II]/H$\alpha$ and [N~II]/H$\alpha$ against H$\alpha$ shows an 
anticorrelation with the line ratios being largest where the H$\alpha$ 
emission is weakest, in agreement with observations.  

We now show some results and 
discuss the effects of varying source luminosity, 
ionizing spectrum, inclusion of extra heating, and ionization by photons 
from leaky H~{\sc ii} regions.  The intensity 
and line ratio maps are qualitatively similar, so our discussion mostly 
focuses on the intensity cuts and scatter plots for the simulations below.

\subsection{Varying Ionizing Luminosity}

In Fig.~6 the ionizing luminosity is varied in the range 
$2\times 10^{49}\,{\rm s}^{-1} \le Q({\rm H}^0) \le 
8\times 10^{49}\,{\rm s}^{-1}$.  
All other parameters are the same as presented in the previous section.  
Increasing or decreasing the source luminosity yields larger and smaller 
ionized volumes.  For the lowest ionizing luminosity, the simulation is 
radiation bounded at $|z|=0.7$~kpc and the vertical cut 
shows that the [O~I]/H$\alpha$ ratio becomes large ($\sim 0.1$) at the 
edge of the ionized volume.  This is due to the combination of high 
temperatures and oxygen rapidly becoming neutral at the edge of the 
ionized volume (see discussion in \S~4.2).  
For lower luminosity 
sources the [S~II]/H$\alpha$ ratio is quite large 
at large $|z|$.  This is because the ionized zone is radiation 
bounded and we are seeing the effects of the interface 
(${\rm S}^{2+}\rightarrow {\rm S}^+$) described above.  Beyond the 
edge of the ionized zone [S~II]/H$\alpha$ is formally infinite because the 
H$\alpha$ intensity is zero.
For this simulation, the resolution of 
our grid is not sufficient to see the rapid decrease in [N~II]/H$\alpha$ 
that occurs in the transition zone where ${\rm N}^{+}\rightarrow {\rm N}^0$.   
We just see the rise of [N~II]/H$\alpha$ and [S~II]/H$\alpha$ with $|z|$ and 
beyond the end of the dotted lines [N~II]/H$\alpha =0$.

\begin{figure}
\psfig{figure=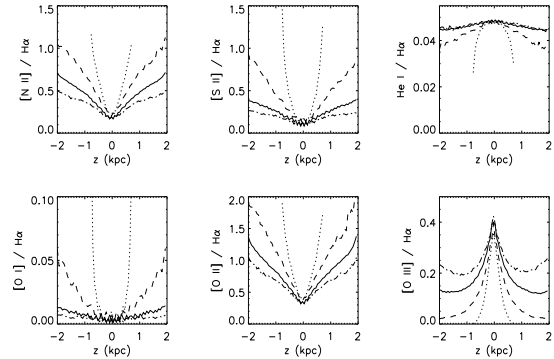,width=3.0truein,angle=0,height=2.0truein}
\caption[]{Vertical cuts at $x=0$ showing the effects of varying ionizing 
luminosity on line ratios.  
Ionizing luminosities are $2\times 10^{49}\,{\rm s}^{-1}$ (dots), 
$4\times 10^{49}\,{\rm s}^{-1}$ (dashed), 
$6\times 10^{49}\,{\rm s}^{-1}$ (solid), and 
$8\times 10^{49}\,{\rm s}^{-1}$ (dot-dashed).  Low luminosities lead to 
a smaller ionized volume so there is a larger relative contribution from 
the ionized to neutral transition zone, hence some line ratios 
are much larger towards large $|z|$.  The lowest luminosity source is 
radiation bounded at $|z|=0.7$~kpc.}
\end{figure}

\subsection{Varying Ionizing Spectra}

The effects of varying the ionizing spectrum are displayed in Figs.~7 and 8.  
Line ratios and scatter plots are shown for WM-basic model 
atmospheres with solar abundances, and 
effective temperatures and $\log g$ values of 
(35000~K, 3.8), (40000~K, 3.75), (45000~K, 3.9), and (50000~K, 4.0).  
The ionizing luminosity is $Q({\rm H}^0) = 6\times 10^{49}\,{\rm s}^{-1}$ 
and all other parameters are as in our standard model.  
As found in other investigations, harder spectra produce higher 
temperatures and increased line ratios for [N~II]/H$\alpha$, 
[S~II]/H$\alpha$, [O~II]/H$\alpha$, and [O~III]/H$\alpha$.  
The 35000~K spectrum has little flux above 24.6~eV 
in the He-ionizing continuum, 
so the He~I/H$\alpha$ ratios are very small for this simulation.

\begin{figure}
\psfig{figure=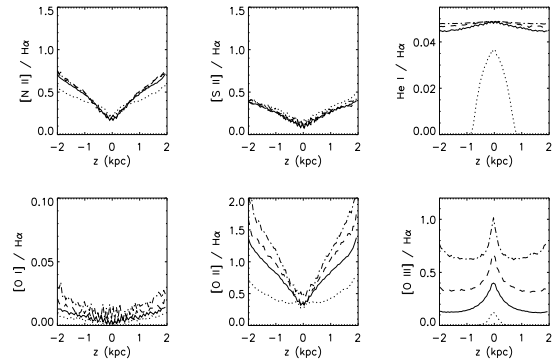,width=3.0truein,angle=0,height=2.0truein}
\caption[]{Effect of varying ionizing spectra on line ratios vertical cuts 
at $x=0$~kpc.  
Effective temperatures for the ionizing source are 35000~K (dots), 
40000K (solid), 45000~K (dashed), and 50000K (dot-dashed). 
The harder ionizing spectra from the 
hotter stars lead to some very large line ratios.}
\end{figure}

\begin{figure}
\psfig{figure=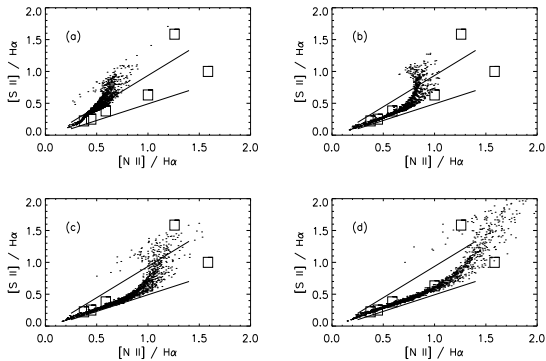,width=3.0truein,angle=0,height=2.0truein}
\caption[]{Effect of varying ionizing spectra on scatter plots.  
Effective temperatures for the ionizing source are (a) 35000~K, 
(b) 40000K, (c) 45000~K, and (d) 50000K.  The harder ionizing spectra from 
the hotter stars lead to increased line ratios, in some cases much larger 
than observed in the local DIG (solid lines) or NGC~891 (squares).}
\end{figure}

\subsection{Composite Models and Extra Heating}

The results of Figs.~7 and 8 indicate that the observed line ratios in the 
local DIG and NGC~891 could be 
reproduced with a range of source spectra and luminosities.  We have 
followed Mathis (2000) and constructed models for a point source with a
composite spectrum that has the following contributions: 56\% from 
$T=35000$~K, 12\% from $T=40000$~K, and the rest from $T=45000$~K.  
This is comparable to the solar neighbourhood where the Garmany, Conti, 
\& Chiosi (1982) catalog shows that around 50\% of the O stars are 
spectral type O8 or later.  
Figure~9 shows that [N~II]/H$\alpha$ and [S~II]/H$\alpha$ increase 
with height above the plane, to values close to those observed in the 
local DIG.  The line ratios are lower than those for NGC~891, likely 
reflecting the hotter sources ionizing its DIG.  As with our other 
simulations, the very large [S~II]/H$\alpha$ line ratios from 
the ionized/neutral interface should be viewed with caution due to 
incomplete physics in our code.

\begin{figure}
\psfig{figure=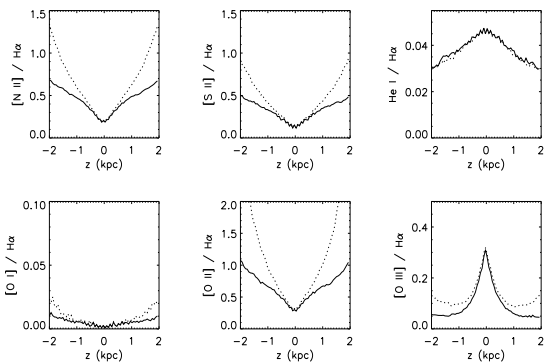,width=3.0truein,angle=0,height=2.0truein}
\caption[]{Composite spectrum models at $x=0$~kpc 
(solid lines) and with additional heating (dots).   }
\end{figure}

\begin{figure}
\psfig{figure=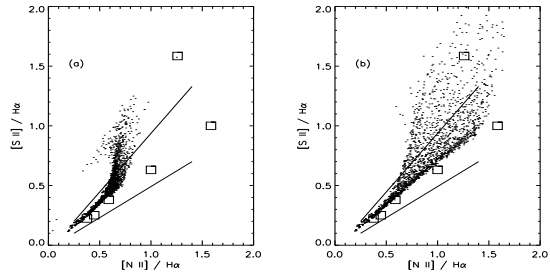,width=3.0truein,angle=0,height=1.5truein}
\caption[]{Composite spectrum scatter plots (left) and with additional 
heating (right) $G_1=5\times 10^{-27} n_e$~ergs/s/cm$^3$.  The lines 
and squares show observations from the local DIG and NGC~891.}
\end{figure}

We have also investigated whether including additional heating may raise 
the [N~II]/H$\alpha$ ratio for this composite model.  
Following Reynolds et al. (1999), we introduce an additional heating term, 
$G_1$, proportional to $n_e$, which will dominate over photoionization 
heating at low $n_e$ at large $|z|$ in our simulations.  The 
models in Fig~9 (dotted curves) and Fig.~10 have 
$G_1=5\times10^{-27} n_e$ ergs/s/cm$^3$, yielding 
temperatures of around 11000~K at $|z|=2$~kpc.  Increasing the heating to 
$G_1= 10^{-26} n_e$ ergs/s/cm$^3$, yields temperatures in excess 
of 15000~K at $|z|=2$~kpc.  The additional heating present in 
Figs.~9 and 10 is much lower than values inferred by Reynolds et al. (1999), 
$G_1\sim 0.65-1.6\times10^{-25} n_e$ ergs/s/cm$^3$
and Mathis (2000),  $\sim 0.75-3\times10^{-25}$ ergs/s/H, 
for 1D models and is 
due to the natural increase of temperatures and line ratios at large $|z|$ 
in our simulations.  Figure~9 shows the [S~II], [N~II], and [O~II] emission 
is boosted compared to the simulation with no additional heating, with 
[O~II]/H$\alpha$ providing a very good probe of additional heating 
(Mathis 2000).  The very high [S~II]/H$\alpha$ line ratios from 
the ionized/neutral interface are still evident as discussed above.

\subsection{Leaky H~{\sc ii} Regions}

\begin{figure}
\psfig{figure=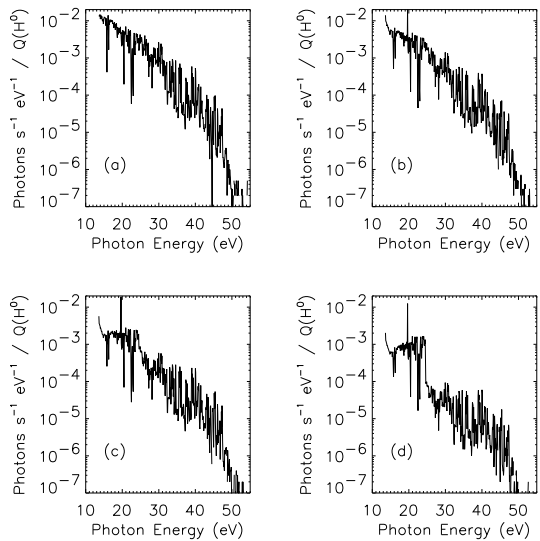,width=3.0truein,angle=0,height=3.0truein}
\caption[]{Input spectra (a) and leaky spectra for leakage fractions 
of (b) 60\%, (c) 30\%, and (d) 15\%.  Note the hardening of the H-ionizing 
photons, the prominent 19.8~eV He~I emission line, and the suppression of 
He-ionizing photons for small escape fractions. }
\end{figure}

\begin{figure}
\psfig{figure=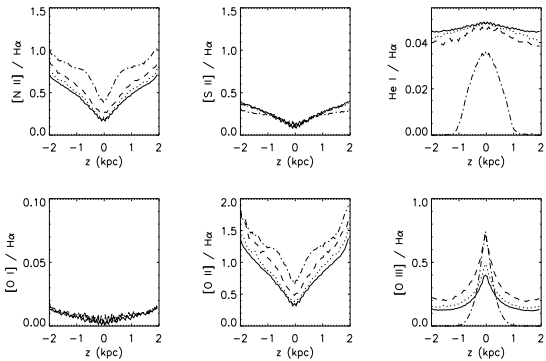,width=3.0truein,angle=0,height=2.0truein}
\caption[]{Models showing the effects of leaky spectra on line 
ratios vertical cuts at $x=0$~kpc.  All models have the same ionizing 
luminosity, but the ionizing spectra are as shown in Fig.~11.  
The lines show a naked star (i.e., 100\% leakage) 40000~K 
model (solid) and leaky models for escape fractions of 60\% (dots), 
30\% (dashed) and 15\% (dot-dashed).}
\end{figure}

\begin{figure}
\psfig{figure=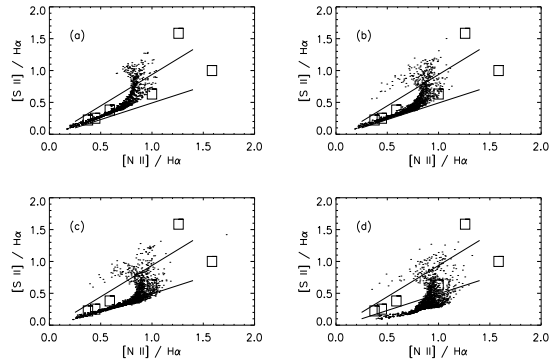,width=3.0truein,angle=0,height=2.0truein}
\caption[]{Effect of leaky spectra models on scatter plots.  
Panels show (a) naked star (i.e., 100\% leakage), and models with escape 
fractions of (b) 60\%, (c) 30\%, and (d) 15\%.}
\end{figure}

Observations indicate that many H~{\sc ii} regions are not 
traditional radiation bounded Str{\"o}mgren spheres, but are density 
bounded and leak a fraction of the source luminosity into the DIG 
(Ferguson et al. 1996; Oey \& Kennicutt 1997; Zurita et al. 2002).  
Depending on the source spectrum for an H~{\sc ii} region, its \
He$^+$ zone may be smaller than the H$^+$ zone (e.g., Osterbrock 1989).  
Therefore, as suggested by Reynolds \& Tufte (1995), 
if an H~{\sc ii} region is density bounded beyond the He$^+$ 
zone, but within the H$^+$ zone, the spectrum leaking into the ISM may 
have its He-ionizing photons suppressed and the H-ionizing continuum 
hardened.  
This is seen in Fig.~4, where the spectrum leaking from a density bounded 
region is harder than the incident spectrum and may have its He-ionizing 
photons suppressed.  Therefore, leaky H~{\sc ii} regions are a very 
plausible way for providing 
higher temperatures, due to the harder H-ionizing continuum, and 
lower He$^+$ fractions, due to the suppression of the He-ionizing photons.  
Hoopes \& Walterbos (2003) presented 1D leaky H~{\sc ii} region models, 
with some models producing elevated [S~II]/H$\alpha$ and [N~II]/H$\alpha$ 
and lower He~I/H$\alpha$.  We generate leaky 
spectra from a constant density H~{\sc ii} region with 
$n({\rm H})=100\,{\rm cm}^{-3}$.  The outer boundary is adjusted until 
the escape fraction is 15\%, 30\%, and 60\% of the incident source spectrum.  
The leaky spectra are then used as the source spectrum for 2D models 
of the DIG.  All DIG models have the same luminosity, 
$6\times 10^{49}$~s$^{-1}$, but differ in the source spectra, shown in 
Fig.~11.

The incident ($T_\star = 40000$~K, $\log g = 3.75$) and escaping 
spectra for our leaky H~{\sc ii} regions are shown in Fig.~11.  
The escaping spectra comprise 
photons that escape directly and contributions from the diffuse radiation 
field: H~I Lyman continuum, He~I Lyman continuum, He~I two-photon continuum, 
and the He~I 19.8~eV, $2^3S\rightarrow 1^1S$ emission line.  The leaky spectra 
are seen to be much harder in the H-ionizing continuum (flatter, or 
even increasing in the interval 13.6~eV -- 24.6~eV) 
and for low 
leakage fractions the He-ionizing continuum is suppressed.  For leaky 
spectra with sources 
hotter than $45000$~K, we found that the He-ionizing continuum is not 
significantly suppressed (see also Hoopes \& Walterbos, Fig.~13).

Figure~12 shows the variation of line ratios with height for the leaky 
spectra of Fig.~11, and Fig.~13 shows the corresponding scatter plots.  
The spectral shape of the leaky spectrum can lead to increased 
or decreased [S~II]/H$\alpha$ and [N~II]/H$\alpha$ (see also 
Hoopes \& Walterbos, Fig.~14).  For the simulations shown here, 
[N~II]/H$\alpha$ increases as the H~{\sc ii} region escape fraction 
decreases (all simulations have the same ionizing luminosity), 
while [S~II]/H$\alpha$ slightly decreases for the lower escape fractions.  
The increase of [N~II]/H$\alpha$ for the lower escape fractions is due to 
the harder spectra between 13.6~eV and 24.6~eV for these models (Fig.~11).  
The He~I/H$\alpha$ decreases for leaky spectra due to absorption of 
He-ionizing photons in the H~{\sc ii} region.  Thus some leaky H~{\sc ii} 
regions provide another way for producing the elevated temperatures, 
line ratios, and low He$^+$ fraction observed in the DIG.

\subsection{Other Models}

We have investigated many more models than those presented above.  Rather 
than present numerous figures, we now describe the main effects of varying 
other parameters.

\subsubsection{Varying Density Structure}

We performed simulations for a single component density structure, 
representing only the DIG component of the ISM.  These simulations had 
lower temperatures, smaller [N~II]/H$\alpha$, [S~II]/H$\alpha$, 
and [O~II]/H$\alpha$, and a larger He~I/H$\alpha$ than the two component 
model.  This was due to the absence of the concentrated density component 
which yields more hardening of the spectrum reaching large $|z|$ in 
the two component density models.  Similarly, the plane parallel models 
presented by Rand (1998) and Bland-Hawthorn et al. (1997) did not produce 
as large [N~II]/H$\alpha$ and [S~II]/H$\alpha$ ratios as our two-component 
models.  It appears that a uniform slab, or single component exponential 
density structure cannot yield the large [N~II]/H$\alpha$, [S~II]/H$\alpha$ 
that are observed without additional heating of the DIG, or hardening of the 
radiation field via leaky H~{\sc ii} region models or a multi-component ISM 
density structure.

\subsubsection{Varying Source Location}

As the source location is moved to larger $z$ heights above the plane, 
the ionized zone becomes radiation bounded for $z<0$ 
(e.g., Miller \& Cox 1993).  At large positive $z$ the grid is more 
ionized since the radiation has less of the concentrated higher density 
component to ionize and pass through before reaching the gas at large $z$.
Compared to a simulation where the source is in the midplane, vertical 
intensity cuts for models where the source is located above the 
midplane show larger line ratios towards the lower radiation bounded edge 
and lower line ratios towards the upper edge of the grid.  These effects 
are seen in the simulations in \S~3.1 where the source luminosity is varied.

\subsubsection{Varying Elemental Abundances}

Using abundances appropriate for the Perseus arm:  (C, N, O, Ne, S)/H = 
(85, 45, 250, 71, 8)~ppm (Mathis 2000) increases the [N~II]/H$\alpha$, 
[S~II]/H$\alpha$, [O~II]/H$\alpha$, and [O~III]/H$\alpha$ line ratios 
by a factor of about 1.5 to 2 compared to those in Fig.~3.  
The lower abundances compared to our standard model result 
in increased line emission to obtain equilibrium temperatures, and hence 
the elevated line ratios.

\subsubsection{Sulfur Dielectronic Recombination}

Accurate modeling of the [S~II] emission is not yet possible due to the 
unknown dielectronic recombination rates for S (see discussion in Ali 
et al. 1991).  In our simulations we used total recombination rates for 
S$^+$ and S$^{2+}$ from Nahar (2000) and the average suggested for S$^{3+}$ 
from Ali et al. (1991).  The rates of Nahar (2000) are from theoretical 
calculations, so may not be as accurate as from experiments.  
We have run our standard simulation with 
no dielectronic recombination for S and also using the averages suggested 
by Ali et al. (1991) for S$^+$, S$^{2+}$, and S$^{3+}$.  We find the 
[S~II]/[N~II] line ratio varies by about a factor of 1.5 among the three 
simulations.  Until more accurate dielectronic 
recombination rates are available, this uncertainty in 
modeling [S~II] emission will remain.

\section{Discussion}

\subsection{Sulfur and Nitrogen}

The model line ratios presented above appear to be in good agreement with 
observations of [S~II] and [N~II] in the local DIG (Haffner et al. 1999), 
NGC~891 (Rand 1998), and several other galaxies 
(e.g., Otte et al. 2001, 2002).  
The increase of [N~II]/H$\alpha$ and [S~II]/H$\alpha$ with $|z|$ is 
a natural consequence of increasing temperatures away 
from the ionizing source due to the hardening of the radiation field.  
Compared to one dimensional averaged models, our two dimensional simulations 
reduce the requirement for extra heating to explain the increasing 
temperatures and line ratios with height above the midplane.

The change of slope in the [S~II]/H$\alpha$ vs [N~II]/H$\alpha$ scatter 
plots (e.g., Fig.~8) are not observed in the Milky Way's DIG 
(Haffner et al. 1999).  The change of slope in our simulations is due to the 
increased S$^+$/S and N$^0$/N fractions at the edge of the ionized volumes.  
The fact that such slope changes are not observed suggests that the DIG 
is almost fully ionized and not density bounded like our single source 
models.  We will investigate multiple source models with overlapping 
ionized regions in a separate paper.  
Alternatively, it is quite likely that our models do not provide 
a good representation of the emission at the ionized/neutral interface as 
we do not include the effects of shocks or ionization fronts.  The role of 
interfaces in interpreting ISM observations is very important (Reynolds 2004), 
and the very large [S~II]/H$\alpha$ and [N~II]/H$\alpha$ line ratios in 
our simulations may be a result of incomplete physics in our simulations.

Some galaxies do show changes in slope of the [S~II]/H$\alpha$ vs 
[N~II]/H$\alpha$ and this has recently been interpreted by 
Elwert, Dettmar, \& T{\"u}llmann (2003) as an indicator for chemical evolution 
in galaxies.  They suggest that increased [S~II]/H$\alpha$ compared to 
[N~II]/H$\alpha$ may arise from younger DIG layers since in the ISM 
nitrogen (from low mass stars, planetary nebulae, and stellar winds) is 
enriched more slowly than sulfur (from spuernovae type~II).  Our models 
adopt uniform abundances throughout the simulation grid and do not address 
this scenario.

\subsection{Oxygen}

There is currently limited data on oxygen lines in the Milky Way's DIG, with 
most observations probing the DIG at $b=0^\circ$.  However, in NGC~891 
there are detailed observations of the dependence of [O~III]/H$\alpha$ 
and [O~I]/H$\alpha$ with height above the plane and Otte et al. (2001, 2002) 
have made measurements of [O~II]/H$\alpha$ in several galaxies.  
Rand (1998) finds that 
[O~III]/H$\alpha$ increases with height, which is opposite to what is 
seen in almost all of our simulations.  The increase of 
[O~III]/H$\alpha$ with $|z|$ in NGC~891 is cited as evidence for O being 
ionized by a different mechanism, such as shocks (e.g., Collins \& Rand 2001), 
instead of pure photoionization.  Our models do not consider dynamics 
or ionization fronts, so cannot address these effects.  
Note, however, that models with very hard spectra 
($T_\star = 50000$~K, Fig.~7) can produce increasing 
[O~III]/H$\alpha$ at large $|z|$.  

Otte et al. (2001, 2002) observe [O~II]/H$\alpha$ and [O~III]/H$\alpha$ 
to increase with height above the plane in five galaxies they studied, 
finding $0.5\la {\rm [O~II]/H}\alpha\la 5$.  
They also observed increases of [S~II]/H$\alpha$ and  
[N~II]/H$\alpha$ with $|z|$.  Our pure photoionization models predict 
$0.5\la {\rm [O~II]/H}\alpha\la 3$, and additional heating can raise 
this even further (e.g., Fig.~9).   It appears that multi-dimensional pure photoionization models can reproduce most of the [O~II]/H$\alpha$ 
observations of Otte et al. (2001, 2002), though additional heating or a 
harder ionizing spectrum may be required for some of the largest 
[O~II]/H$\alpha$ ratios.

The [O~I]/H$\alpha$ line ratios in our models show that 
[O~I] increases in strength towards the edge of the ionized zone.  The 
strong charge-exchange coupling of O$^0$ to H$^0$ and the increased 
temperatures towards the edge of the ionized zone result in the 
increased [O~I]/H$\alpha$.  This is generally seen as a problem with 
photoionization models, since in the few observations in the local DIG, 
albeit at $b=0^\circ$, [O~I] is observed to be rather weak with 
[O~I]/H$\alpha \la 0.03$.  The [O~I] emission may be reduced if the 
region is fully ionized, or density bounded instead of radiation bounded 
(e.g., Mathis 2000; Sembach et al. 2000).  This is seen in the vertical 
cuts showing that [O~I]/H$\alpha < 0.03$ in models where the gas 
is ionized beyond the maximum $|z|$ of our simulation box (e.g., Fig.~3).  
The increasing [O~I]/H$\alpha$ with $|z|$ in some of our simulations 
(e.g., Fig.~6) do 
match the data for NGC~891, where [O~I]/H$\alpha\sim 0.1$ at large $|z|$ 
(Rand 1998).  More observations of [O~I] at larger $|z|$ in the Galactic 
DIG will determine whether there really is a difference in the [O~I] 
emission between the Milky Way and NGC~891.

\subsection{Helium}

As with [O~I], the few measurements of He~I in the Galactic DIG are 
at $b=0^\circ$.  The observations indicate that helium is 
underionized relative to hydrogen with ${\rm He}^+/{\rm H}^+ \la 0.04$ 
(Reynolds \& Tufte 1995; see also Heiles et al. 1996).  
The He~I observations probe the ionizing 
spectrum for the DIG and indicate a relatively soft spectrum, typically 
spectral type O8 or later, corresponding to $T_\star \sim 36000$~K.  
The situation is not as extreme in NGC~891 where He$^+$/H$^+ \sim 0.06$, 
indicating a harder ionizing spectrum for its DIG (Rand 1997).

Our simulations using a composite spectrum, representative of the O stars 
in the solar neighbourhood, produce fairly low values for He~I/H$\alpha$, 
in line with current observations at $b=0^\circ$.  Further observations 
of He~I at higher latitudes will provide additional tests of our models 
and the hardening of the radiation field at large $|z|$.

We have also investigated an alternative mechanism for reducing the 
helium ionization, even when the sources are hotter than O8.  
The basic idea is that 
the radiation leaking out of midplane H~{\sc ii} regions to ionize 
the DIG may have its helium ionizing photons suppressed (e.g., 
Hoopes \& Walterbos 2003).  
Our leaky H~{\sc ii} region models do indeed show that low values for 
He$^+$/H$^+$ may be obtained, even if the ionizing source is quite hot.  
For the same ionizing luminosity, these models also produce higher 
temperatures than our other models, since the leaking ionizing spectrum is 
harder in the H-ionizing continuum.  
Therefore, leaky H~{\sc ii} regions provide a plausible mechanism for 
explaining the low helium ionization in the Galactic DIG.

\section{Summary}

We have presented some of the first multi-dimensional 
photoionization simulations investigating intensity and line ratio maps 
in models of the Milky Way's diffuse ionized gas.  Our models 
produce [S~II]/H$\alpha$ and [N~II]/H$\alpha$ line ratios that are in good 
agreement with observations.  Our simulations 
reproduce the observed increase of [S~II]/H$\alpha$ and [N~II]/H$\alpha$ 
with increasing $|z|$ due to the increasing temperatures at large 
distances from the ionizing sources.  Previous analyses of line 
ratios in the DIG suggested additional heating was required to 
reproduce the elevated line ratios at large $|z|$.  However, this 
appears to be due to the use of spherically averaged photoionization 
models.  

Most of our simulations do not include any additional 
heating above that provided by pure photoionization.  When we did include 
extra heating it was much less than used in 1D photoionization models 
of the DIG.  Models such as those presented here, where line ratios 
are constructed for rays piercing the outer edges of photoionized 
regions, are more appropriate for the DIG.  
We have also shown that leaky H~{\sc ii} regions provide a 
plausible explanation for the low helium ionization in the DIG.  The 
hardened leaky spectra produce higher temperatures than our standard 
models, in agreement with other work (Hoopes \& Walterbos 2003).

Unless we use very hard ionizing spectra, our models do not reproduce 
the observed increase of [O~III]/H$\alpha$ 
with $|z|$ in NGC~891.  The observed rise of [O~III]/H$\alpha$ may be due 
to hotter ionizing sources or an additional source 
of ionization not present in our photoionization 
simulations (e.g., shocks or massive stars formed at large $|z|$ from 
the dense clouds observed there, Howk \& Savage 1997).  
Another potential problem with our simulations is that 
they do not treat the physics at the ionization front, possibly resulting 
in an overprediction of the [S~II]/H$\alpha$ line 
ratios.  Despite these shortcomings, our models 
clearly demonstrate the importance 
of geometry in models of the DIG: compared to homogeneous or 
single component densities, multi-component models produce 
more hardening of the radiation field, and more elevated temperatures 
at large $|z|$.  
In a future paper we will explore 
3D models incorporating the known locations, luminosities, and spectral 
types of O stars in the Solar neighbourhood.  The WHAM data in addition 
to the H~I atlas (Hartmann \& Burton 1997) 
should enable us to place constraints on the ISM density structure.  

We thank Alison Campbell, Torsten Elwert, Barbara Ercolano, Matt Haffner, 
Kirk Korista, Lynn Matthews, 
John Raymond, Ron Reynolds, Jon Slavin, and Steve Tufte for many useful 
discussions relating to this work.  Suggestions from an anonymous referee 
resulted in a clearer presentation of our results.  
KW acknowledges support from a PPARC Advanced Fellowship, JSM claims to 
be retired.

\label{lastpage}


\begin{thebibliography}{}

\bibitem[\protect\citeauthoryear{Ali et al.}{1991}]{ali91}
Ali, B., Blum, R.D., Bumgardner, T.E., Cranmer, S.R., Ferland, G.J., 
Haefner, R.I., \& Tiede G.P. 1991, PASP, 103, 1182

\bibitem[\protect\citeauthoryear{Anders \& Grevesse}{1989}]{ag89}
Anders, E., \& Grevesse, N. 1989, Geochim, Cosmochim. Acta, 53, 197

\bibitem[\protect\citeauthoryear{Bland-Hawthorn et al.}{1997}]{bfq97}
Bland-Hawthorn, J., Freeman, K.C., \& Quinn, P.J. 1997, ApJ, 143, 155

\bibitem[\protect\citeauthoryear{Ciardi et al.}{2002}]{c02}Ciardi, 
B., Bianchi, S., \& Ferrara, A. 2002, MNRAS, 331, 463

\bibitem[\protect\citeauthoryear{Collins \& Rand}{2001}]{cr01}
Collins, J.A., \& Rand, R.J. 2001, ApJ, 551, 57

\bibitem[\protect\citeauthoryear{Dickey \& Lockman}{1990}]{dl90}
Dickey, J.M., \& Lockman, F.J. 1990, ARA\&A, 28, 215

\bibitem[\protect\citeauthoryear{Domgoergen \& Dettmar}{1997}]{dd97}
Domgoergen, H., \& Dettmar, R.-J. 1997, A\&A, 322, 391

\bibitem[\protect\citeauthoryear{Domgoergen \& Mathis}{1994}]{dm94}
Domgoergen, H., \& Mathis, J.S. 1994, ApJ, ApJ, 428, 647

\bibitem[\protect\citeauthoryear{Dove \& Shull}{1994}]{ds94}Dove,
J.B., \& Shull, J. M. 1994, ApJ, 423, 196

\bibitem[\protect\citeauthoryear{Elwert et al.}{2003}]{edt03}
Elwert, T., Dettmar, R.-J., \& T{\"u}llmann, R. 2003, BAAS, 203, 111.05

\bibitem[\protect\citeauthoryear{Ercolano et al.}{2003}]{erc}
Ercolano, B., Barlow, M. J., Storey, P. J., \& Liu, X.-W. 2003, MNRAS,
340, 1136

\bibitem[\protect\citeauthoryear{Ferguson et al.}{1996}]{f96}
Ferguson, A.M., Wyse, R.F.G., Gallagher, J.S., \& Hunter, D.A. 
1996, AJ, 111, 2265

\bibitem[\protect\citeauthoryear{Garmany et al.}{1982}]{gcc82}
Garmany, C.D., Conti, P.S., \& Chiosi, C. 1982, ApJ, 263, 777

\bibitem[\protect\citeauthoryear{Haffner et al.}{1999}]{hrt99}
Haffner, L.M., Reynolds, R.J., \& Tufte, S.L. 1999, ApJ, 523, 223

\bibitem[\protect\citeauthoryear{Hartmann \& Burton}{1997}]{hb97}
Hartmann, D., \& Burton, W.B. 1997, ``Atlas of Galactic Neutral Hydrogen,'' 
Cambridge University Press

\bibitem[\protect\citeauthoryear{Heiles et al.}{1996}]{h96}
Heiles, C., Koo, B-C., Levenson, N. A. \&  Reach, W. T. 1996, 
ApJ, 462, 326

\bibitem[\protect\citeauthoryear{Hoopes et al.}{1996}]{hwg96}
Hoopes, C.G., Walterbos, R.A.M., \& Greenawalt, B.E. 1996, AJ, 112, 1429

\bibitem[\protect\citeauthoryear{Hoopes \& Walterbos}{2003}]{hw03}
Hoopes, C.G., \& Walterbos, R.A.M. 2003, ApJ, 586, 902

\bibitem[\protect\citeauthoryear{Howk \& Savage}{1997}]{hs97}
Howk, J.C., \& Savage, B.D. 1997, AJ, 114, 2463

\bibitem[\protect\citeauthoryear{Mathis}{2000}]{m00}Mathis, J. S. 2000,
ApJ, 544, 347

\bibitem[\protect\citeauthoryear{Mathis}{1986}]{m86}
Mathis, J. S. 1986, ApJ, 301, 423

\bibitem[\protect\citeauthoryear{Miller \& Cox}{1993}]{mc}
Miller, W. W., III, \& Cox, D. P.1993, ApJ, 417, 579

\bibitem[\protect\citeauthoryear{Minter \& Spangler}{1997}]{ms} 
Minter, A. H., \& Spangler, S. R., 1997, ApJ, 458, 194

\bibitem[\protect\citeauthoryear{Nahar}{2000}]{n2000} 
Nahar, S.N. 2000, ApJS, 126, 537

\bibitem[\protect\citeauthoryear{Och, Lucy, \& Rosa}{1998}]{olr}
Och, S. R., Lucy, L. B., \& Rosa, M. R. 1998, A\&A, 336, 301

\bibitem[\protect\citeauthoryear{Oey \& Kennicutt}{1997}]{ok97}
Oey, M.S., \& Kennicutt, R.C. 1997, MNRAS, 291, 827

\bibitem[\protect\citeauthoryear{Otte et al. 2002}{2002}]{o2002}
Otte, B., Gallagher, J.S., \& Reynolds, R.J. 2002, ApJ, 572, 823

\bibitem[\protect\citeauthoryear{Otte et al. 2001}{2001}]{o2001}
Otte, B., Reynolds, R.J., Gallagher, J.S., \& Ferguson, A.M.N. 2001, 
ApJ, 560, 207

\bibitem[\protect\citeauthoryear{Otte \& Dettmar}{1999}]{od99}
Otte, B., \& Dettmar, R.-J. 1999, A\&A, 343, 705

\bibitem[\protect\citeauthoryear{Osterbrock}{1989}]{o89}Osterbrock,
D. E. 1989, Astrophysics of Gaseous Nebulae and Active Galactic Nuclei,
University Science Books, Mill Valley, CA

\bibitem[\protect\citeauthoryear{Pauldrach et al.}{2001}]{p01}
Pauldrach, A.W.A., Hoffmann, T.L., \& Lennon, M. 2001, A\&A, 375, 161

\bibitem[\protect\citeauthoryear{Rand}{1998}]{rand98}
Rand, R. 1998, ApJ, 501, 137

\bibitem[\protect\citeauthoryear{Rand}{1997}]{rand97}
Rand, R. 1997, ApJ, 474, 129

\bibitem[\protect\citeauthoryear{Raymond}{1992}]{raymond92}
Raymond, J.C. 1992, ApJ, 384, 502

\bibitem[\protect\citeauthoryear{Reynolds}{2004}]{ron04}
Reynolds, R.J. 2004, to appear in ``How Does the Galaxy 
Work?'', E.J. Alfaro, E. Perez, \& J. Franco, eds. 

\bibitem[\protect\citeauthoryear{Reynolds, Haffner, \& Madsen}{2002}]
{r02}Reynolds, R. J., Haffner, L. M., \& Madsen, G. J., 2002, in
``Galaxies: the Third Dimension'', M. Rosada, L. Binette, \& L. Arias,
eds. (Astr. Soc. Pacific,: San Francisco)

\bibitem[\protect\citeauthoryear{Reynolds, Haffner, \& Tufte}{1999}]
{rht}Reynolds, R. J., Haffner, L. M., \& Tufte, S.L. 1999, ApJ, 525,
L21

\bibitem[\protect\citeauthoryear{Reynolds \& Tufte}{1995}]{rt}
Reynolds, R. J., \& Tufte, S. L. 1995, ApJ, 448, 715

\bibitem[\protect\citeauthoryear{Reynolds \& Cox}{1992}]{rc92}
Reynolds, R. J., \& Cox, D.P. 1992, ApJ, 400, L33

\bibitem[\protect\citeauthoryear{Reynolds}{1985}]{r85}
Reynolds, R. J. 1985, ApJ, 294, 256

\bibitem[\protect\citeauthoryear{Sembach et al.}{2000}]{S2000}
Sembach, K.R., Howk, J.C., Ryans, R.S.I., \& Keenan, F.P. 2000, 
ApJ, 528, 310

\bibitem[\protect\citeauthoryear{Slavin et al.}{1993}]{ssb93}
Slavin, J.D., Shull, J.M., \& Begelman, M.C. 1993, ApJ, 407, 83

\bibitem[\protect\citeauthoryear{Sternberg et al.}{2004}]{shp04}
Sternberg, A., Hoffmann, T.L., \& Pauldrach A.W.A., 2004, ApJ, in press, 
astro-ph/0312232

\bibitem[\protect\citeauthoryear{Verner \& Ferland}{1996}]{vf96}
Verner, D. A, \& Ferland, G. J. 1996, ApJS, 103, 467

\bibitem[\protect\citeauthoryear{Wang et al.}{1998}]{whl98}
Wang, J., Heckman, T.M., \& Lehnert, M.D. 1998, ApJ, 509, 93

\bibitem[\protect\citeauthoryear{Wood et al.}{2004}]{wme04}
Wood, K., Mathis, J.S., \& Ercolano, B. 2004, MNRAS, in press, 
astro-ph/0311584

\bibitem[\protect\citeauthoryear{Wood \& Loeb}{2000}]{wl}
Wood, K., \& Loeb, A. 2000, ApJ, 545, 86

\bibitem[\protect\citeauthoryear{Zurita et al.}{2002}]{z02}
Zurita, A., Beckman, J.E., Rozas, M., \& Ryder, S. 2002, 
A\&A, 386, 801

\end{thebibliography}
\end{document}